\documentclass[amsmath, amssymb, superscriptaddress, reprint, twocolumn, colorlinks=true, citecolor=blue, linkcolor=blue, urlcolor=blue]{revtex4-2}
\let\makeuppercase\MakeUppercase
\usepackage[utf8]{inputenc}
\usepackage[T1]{fontenc}
\usepackage{graphicx}
\usepackage{grffile}
\usepackage{longtable}
\usepackage{wrapfig}
\usepackage{rotating}
\usepackage[normalem]{ulem}
\usepackage{amsmath}
\usepackage{textcomp}
\usepackage{amssymb}
\usepackage{capt-of}
\usepackage{hyperref}
\usepackage{mathptmx}
\usepackage{upgreek}
\usepackage{float}
\usepackage{bm}
\usepackage{setspace}
\usepackage{booktabs}
\usepackage{subfigure}
\usepackage{url}
\usepackage{xcolor}
\usepackage{multirow}
\usepackage{array}
\usepackage{tabularx}
\usepackage{nicematrix}
\usepackage{blkarray}
\usepackage{mathptmx}
\usepackage{mathrsfs}
\usepackage{xcolor}
\usepackage{diagbox}
\makeatletter
\newcommand{\coloredcite}[1]{{\color{ccr}\cite{#1}}}

\makeatother

\definecolor{ccr}{RGB}{45,47,145} 

\hypersetup{colorlinks=true, citecolor=ccr, urlcolor=ccr, linkcolor=ccr}

\DeclareMathAlphabet{\mathcal}{OMS}{cmsy}{m}{n}

\date{\today}
\UseRawInputEncoding
\begin{document}

\title{Charge Transport and Mode Transition in Dual-Energy Electron Beam Diodes}

\author{Chubin Lin}
\affiliation{Department of Electrical Engineering, Tsinghua University, Beijing 100084, China}

\author{Jiandong Chen}
\affiliation{Department of Electrical Engineering, Tsinghua University, Beijing 100084, China}

\author{Huihui Wang}
\affiliation{Department of Chemical Engineering, Tsinghua University, Beijing 100084, China}

\author{Yangyang Fu}
\email{fuyangyang@tsinghua.edu.cn}
\affiliation{Department of Electrical Engineering, Tsinghua University, Beijing 100084, China}
\affiliation{State Key Laboratory of Power System Operation and Control, Department of Electrical Engineering, Tsinghua University, Beijing 100084, China}
\affiliation{Sichuan Energy Internet Research Institute, Tsinghua University, Sichuan 610213, China}

\begin{abstract}
This Letter uncovers five distinct charge transport modes and their transitions in dual-energy electron beam diodes.
We via first-principle particle-in-cell (PIC) simulations establish that the specific mode (e.g., space charge oscillations) and the current transmitted characteristics are essentially governed by the interplay between the electron beam energy and injected current density.
The effective space charge limited (SCL) current (lower than conventional SCL threshold) is unveiled due to the coupled interactions between two electron beams.
A generalized analysis is conducted for \(n\)--component electron beams, and a theoretical piecewise function is for the transmitted current density proposed, which agrees well with the PIC results under designed conditions.
The discovery provides a mechanistic picture of multiple electron beam transport in diodes, paving the way for novel designs of high-performance modern vacuum electronic devices.
\end{abstract}

\maketitle

\textit{Introduction}---Electron beams \coloredcite{Tsimring_2007,Go_2022,Sharma_2025}, as directed electron streams characterized by their current density \coloredcite{Li_2025,Campanell_2012,Campanell_2016}, energy spectrum \coloredcite{Wenz_2019,Furman_2002,chen_2025_PRE} and spatiotemporal evolutions \coloredcite{Heri_2025,Ang_2007}, are indispensable drivers for various applications \coloredcite{Zhang_2017,Munoz_2024,Go_2024,chen_2025_PRAppl}, such as Pierce diodes \coloredcite{Ender_2000}, virtual cathode oscillators \coloredcite{Jiang_2024,Mumtaz_2024}, thermionic energy converters \coloredcite{Sitek_2021,Khalid_2016,Lin_2024}, compact terahertz sources \coloredcite{Chen_2026,Alexandersson_2026,Andreas_2010,Zhang_2025,Kumar_2025}, and ultrafast switches \coloredcite{Nikoo_2020}.
In electron beam driven diodes, an injected intense electron current, accompanied by significant space charge effects \coloredcite{Birdsall_1966,Campanell_2025}, can limit the transmitted current to a maximum, which is known as the space-charge-limited (SCL) current \(J_\text{SCL}\) \coloredcite{Child_1911,Jaffe_1944}.
The charge transport in diodes with mono-energetic electron beams has been well understood, and the maximum transmitted current obeys the prediction by the SCL current \(J_\text{SCL}\) \coloredcite{Zhu_2013,Garner_2023,Garner_2025}.
Previously, first-principle particle-in-cell (PIC) simulations showed that in mono-energetic electron beam driven diodes, the transmitted current \(J_\text{tran}\) is as a function of the injected current \(J_\text{inj}\), i.e., \(J_\text{tran}=J_\text{inj}\) for \(J_\text{inj} \leq J_\text{SCL}\), whereas \(J_\text{tran}\) becomes saturated (or oscillating with nonzero emission velocity) for \(J_\text{inj}>J_\text{SCL}\) \coloredcite{Lafleur_2020,Lafleur_2022}.
If \(J_\text{inj}>J_\text{SCL}\), a virtual cathode (adjacent to the cathode) can form with an oscillating potential barrier, which periodically reflects less energetic electrons released from the cathode, thereby inducing space-charge oscillations \coloredcite{Bridges_1963,Lin_2025,Zhu_2023}.
In the oscillation regime, existing analytical solutions can predict the spatial profiles of diode physical quantities (e.g., electric potential) \coloredcite{Lafleur_2020}.
The scaling law for the space charge oscillation frequency has been theoretically derived and validated via PIC simulations \coloredcite{Lin_2025_pre}.
However, the SCL current fails when electrons have distributed energies (e.g., the Maxwell distribution for electron thermal motion) or when multi-energy electron beams form, since space-charge oscillations can be altered, leading to different steady states \coloredcite{Lafleur_2020}.
Although well-studied in mono-energy beams, the SCL current and oscillatory dynamics in diodes with discrete energy components present emergent complexities, and a complete theoretical framework remains elusive.

In this Letter, we report five complete charge transport modes and their transitions in diodes driven by precisely controlled dual-energy electron beams.
For the first time, we, via PIC simulations, demonstrate the control of five distinct operating modes by independently modulating high- and low-energy electron beams, and identify that the velocity ratio dictates mode transitions, whereas the injected current ratio determines transmitted current scaling characteristics. 
The results provide key insights into multi-electron beam-driven diodes, which are essential for promoting modern technologies in vacuum electronics.

\begin{figure}[htbp]
\centering
\includegraphics[clip, width=0.95\linewidth]{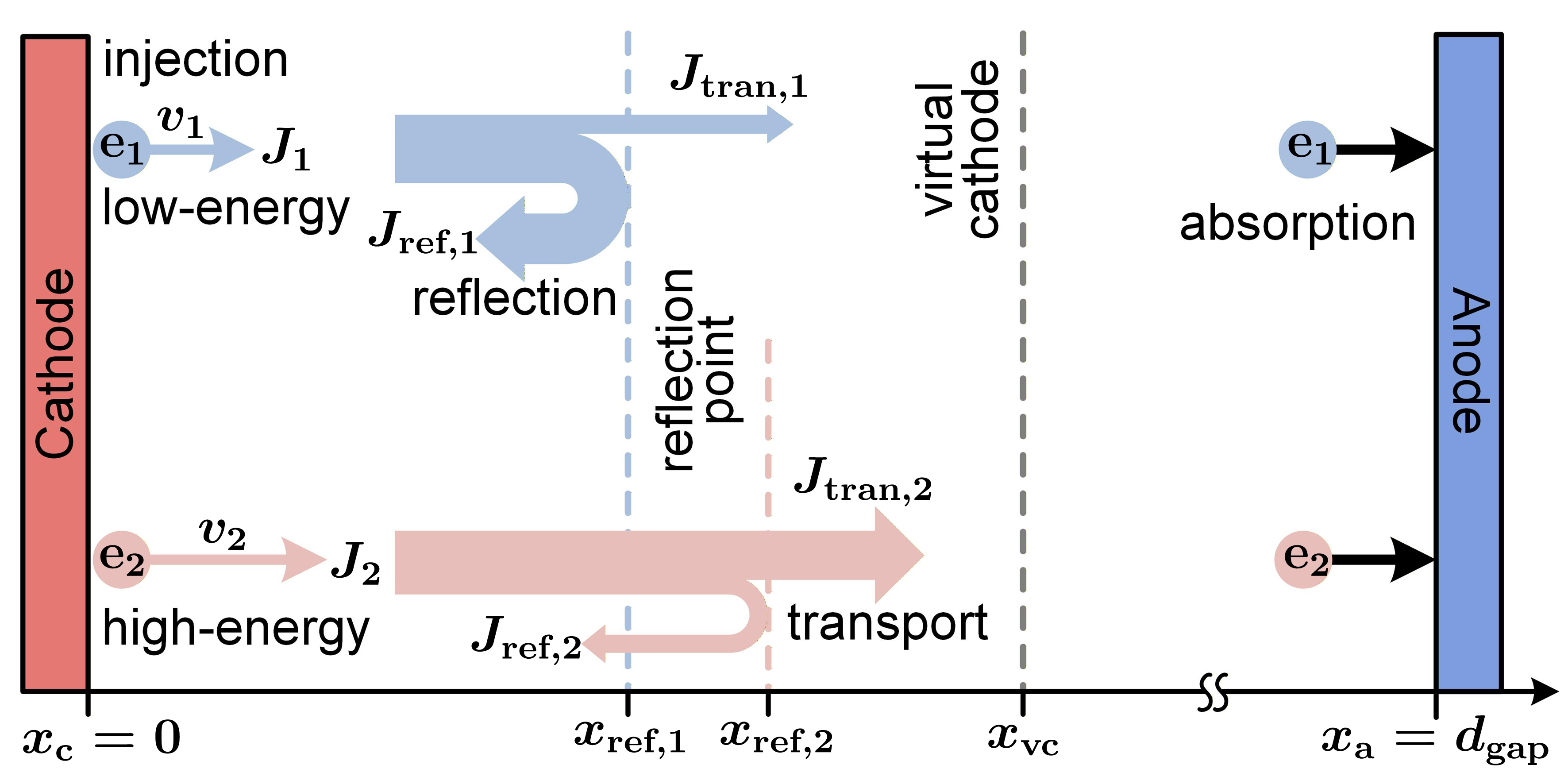}
\caption{\label{fig:Fig0}
Schematic of the dual-energy electron beam driven diode, in which the cathode injection current consists of low-energy and high-energy electrons. 
Low-energy electrons tend to be reflected mostly, whereas high-energy electrons can be transmitted predominantly to the anode.
\(x_\text{c}=0\) is the cathode, and \(x_\text{a}=d_\text{gap}\) is the anode.
Note that the cathode-to-virtual cathode region is enlarged and \(x_\text{ref,1}<x_\text{ref,2}<x_\text{vc}\ll d_\text{gap}\), where \(x_\text{ref,1}\) and \(x_\text{ref,2}\) are the reflection points, and \(x_\text{vc}\) is the virtual cathode position.}
\end{figure}

\textit{Physical model}---The schematic of the planar vacuum diode with cathode-anode gap distance of \(d_\text{gap}\) driven by dual-energy electron beams is shown in Fig.~\ref{fig:Fig0}.
Two groups of electrons are injected from the cathode, including low-energy electrons \(\text{e}_1\) at an initial velocity \(v_1=\beta_1\sqrt{2eV/m_{e}}\) with an injected current density \(J_1\) and high-energy electrons \(\text{e}_2\) at an initial velocity \(v_2=\beta_2\sqrt{2eV/m_e}\) with an injected current density \(J_2\), where \(\beta_1\) and \(\beta_2\) are dimensionless initial velocity factors, \(e\) is the elementary charge, \(V\) is the gap voltage, and \(m_{e}\) is the electron mass.
The different groups of electron beams, generated from field emission \coloredcite{Frobes_2019,Ying_2025}, thermal-field emission \coloredcite{Jensen_2019}, photo-emission \coloredcite{Heimerl_2024}, or laser-driven accelerators \coloredcite{Wenz_2019}, can be independently transported.
The total injected current density \(J_0\) is the sum of injected current density of \(\text{e}_1\) and \(\text{e}_2\), i.e., \(J_0=J_1+J_2\).
Here, the initial velocity and injected current of each electron beam can be independently controlled, which are characterized by two dimensionless parameters, i.e., the velocity ratio \(v_1/v_2=\beta_1/\beta_2 \in (0,1)\) and the injected current ratio \(J_1/J_2\) (can be controlled greater or less than one).

The virtual cathode with the minimum potential \(\phi_\text{vc}\) at the position \(x_\text{vc}\) is generated self-consistently by both electron beams, which can block electrons when their initial energy \(\mathcal{E}_0<-e\phi_\text{vc}\). 
Here we have \(\mathcal{E}_0=\mathcal{E}_{e1}\) for \(\text{e}_1\), \(\mathcal{E}_0=\mathcal{E}_{e2}\) for \(\text{e}_2\), and \(\mathcal{E}_{e1}<\mathcal{E}_{e2}\), thus, electrons with different energies exhibit distinct dynamical behaviors responding to the electric potential. 
As shown in Fig.~\ref{fig:Fig0}, low-energy electrons \(\text{e}_1\) without sufficient kinetic energy can be predominantly reflected by the virtual cathode potential barrier, whereas high-energy electrons \(\text{e}_2\) can overcome potential barriers, and can be mostly transported and absorbed by the anode.
The electron reflection point is determined by the critical condition where the reflection potential \(\phi_\text{ref}=-\mathcal{E}_0/e\), and here we have \(\phi_\text{ref,1}=-\mathcal{E}_{e1}/e\) for \(\text{e}_1\) and \(\phi_\text{ref,2}=-\mathcal{E}_{e2}/e\) for \(\text{e}_2\). Thus, the electron reflection point position for low-energy electrons is closer to the cathode than for high-energy electrons, i.e., \(x_\text{ref,1} < x_\text{ref,2}\).

The PIC simulations are conducted using 1D VSim\(^{\textcircled{c}}\) PIC code \cite{Vsim}. The gap distance ($d_\text{gap}=100\ \mu\text{m}$) is discretized by 4000 uniform cells ($\Delta x = 25\ \text{nm}$) with a time step of $\Delta t = 1\ \text{fs}$, following the Courant-Friedrichs-Lewy condition. 
Dual-energy electron beams with respective fluxes and energies (zero temperature) are injected from the cathode surface into vacuum, with 100 macroparticles injected into the first grid (cathode-vacuum side) per time step.
For boundary conditions, the electrons can be fully absorbed at the anode and cathode.
Considering the statistical noise of the PIC simulations and the fluctuations caused by periodic oscillations, the results of the collected current at the electrodes are time-averaged over multiple periods \cite{Lafleur_2020}.

\textit{Operating modes}---Each electron beam (low-energy \(\text{e}_1\) and high-energy \(\text{e}_2\)) exhibits three possible charge transport states: fully transported mode (T--mode), oscillation mode (O--mode) with intermittent transport and reflection, and fully reflected mode (R--mode).
While state permutations permit \(3^2=9\) theoretical transport modes, stronger space-charge suppression on \(\text{e}_1\) than \(\text{e}_2\) constrains the system to five dynamical modes (M1--M5). Notably, both \(\text{e}_1\) and \(\text{e}_2\) cannot be fully reflected simultaneously; otherwise, the virtual cathode is unable to sustain \coloredcite{Lafleur_2020}. 
By systemically tuning the control parameters of each electron beam, e.g., initial velocity and injected current density, we identify five observationally distinct operating modes in dual-energy electron beam driven diodes, as shown in Table~\ref{Table1}.
The "NA" indicates the combination does not exist.

\begin{table}[htbp]
\caption{\label{Table1}Operating modes in dual-energy electron-beam-driven diodes, in which \(\text{e}_1\) and \(\text{e}_2\) are at low and high energies, respectively.}

\renewcommand{\arraystretch}{1} 
\begin{ruledtabular}
\begin{tabular}{cccc}
\diagbox[width=1.0cm]{\(\text{e}_2\)}{\(\text{e}_1\)} & \text{T} & \text{O} & \text{R}  \\
\hline
  \text{T} & \text{M1 (\(\text{T}_1-\text{T}_2\))} & \text{M2 (\(\text{O}_1-\text{T}_2\))} & \text{M3 (\(\text{R}_1-\text{T}_2\))}\\
 \text{O} & \text{NA} & \text{M5 (\(\text{O}_1-\text{O}_2\))} & \text{M4 (\(\text{R}_1-\text{O}_2\))}\\ 
        \text{R} & \text{NA} & \text{NA} & \text{NA}\\      
\end{tabular}
\end{ruledtabular}
\footnotetext[1]{Default journal substyle.}
\end{table}

\begin{figure}[b]
\centering
\includegraphics[clip, width=0.84\linewidth]{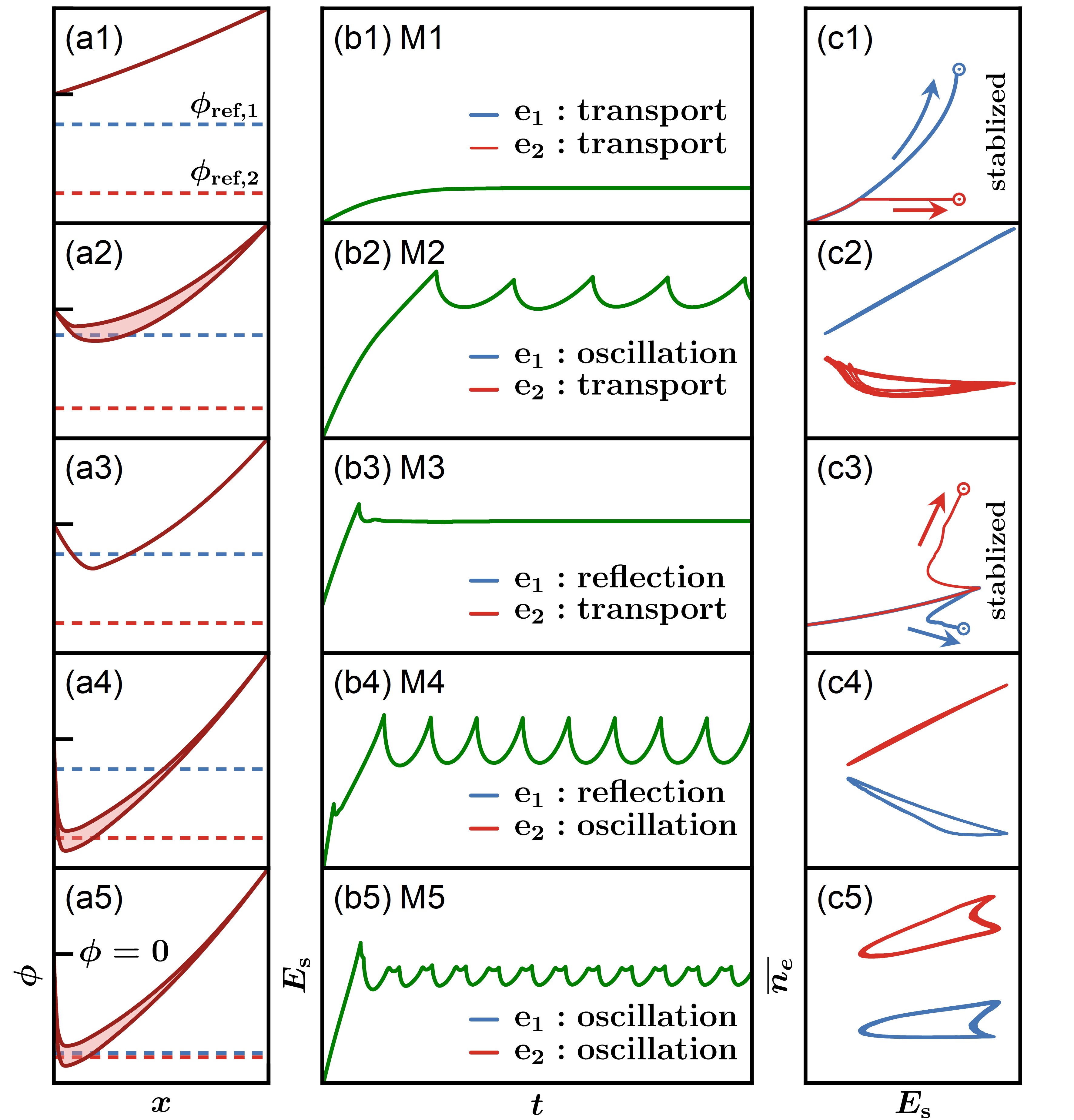}
\caption{\label{fig:Fig1}
Five operating modes in the dual-energy electron-beam-driven diode. The panels from left to right correspond to (a1)--(a5) \(\phi\)--\(x\), (b1)--(b5) \(E_\text{s}\)--\(t\), and (c1)--(c5) \(\bar{n}_e\)--\(E_\text{s}\).
The combinations of the emitted electrons (e\(_1\) and e\(_2\)) result in five operating modes, i.e., M1 (both e\(_1\) and e\(_2\) transmitted), M2 (e\(_1\) oscillated and e\(_2\) transmitted), M3 (e\(_1\) reflected and e\(_2\) transmitted), M4 (e\(_1\) reflected and e\(_2\) oscillated), and M5 (both e\(_1\) and e\(_2\) oscillated).}
\end{figure}

Figure~\ref{fig:Fig1} demonstrates five operating modes (M1--M5), including the electric potential distribution \(\phi-x\), temporal evolution of the cathode surface electric field \(E_\text{s}-t\), and phase space of spatially averaged electron density versus cathode surface electric field \(\overline{n}_e-E_\text{s}\) for each mode.
In mode M1, the minimum gap potential follows \(\phi_\text{min}>\phi_\text{ref,1}>\phi_\text{ref,2}\) [see Fig.~\ref{fig:Fig1}(a1)], which enables fully transport of both low-energy electrons \(\text{e}_1\) and high-energy electrons \(\text{e}_2\) in steady state.
Mode M2 emerges when the intense space-charge effect occurs and lowers the electric potential, forming a virtual cathode potential \(\phi_\text{vc}\) that oscillates near the reflection potential of low-energy electrons \(\phi_\text{ref,1}\).
This condition, though maintain full transport for \(\text{e}_2\), can trigger intermittent reflection-transport cycles for \(\text{e}_1\), which can induce oscillations of physical quantities, such as the cathode surface electric field \(E_\text{s}\) [Fig.~\ref{fig:Fig1}(b2)].
Since the oscillation is induced by \(\text{e}_1\), the oscillation frequency for M2 can be predicted by \(f_\text{os}\varpropto\sqrt{J_1/\beta_1}\) \coloredcite{Lin_2025_pre}.
Mode M3 is established when the virtual cathode potential stabilizes below \(\phi_\text{ref,1}\) but above \(\phi_\text{ref,2}\), i.e., \(\phi_\text{ref,1}>\phi_\text{min}>\phi_\text{ref,2}\), [Fig.~\ref{fig:Fig1}(a3)], resulting in a new steady state where \(\text{e}_1\) is completely reflected but \(\text{e}_2\) is fully transported.
In Fig.~\ref{fig:Fig1}(a4), by further increasing the injected current, M4 occurs when the virtual cathode potential is sufficiently low to oscillate near the reflection potential of high-energy electrons \(\phi_\text{ref,2}\).
Thus, \(\text{e}_1\) can be completely reflected and reabsorbed by the cathode whereas \(\text{e}_2\) has intermittent transport and reflection, inducing periodic oscillation with the oscillation frequency scaling as \(f_\text{os}\varpropto\sqrt{J_2/\beta_2}\).
Especially, when the energies of \(\text{e}_1\) and \(\text{e}_2\) are rather close (e.g., \(v_1/v_2=\sqrt{\mathcal{E}_{e1}/\mathcal{E}_{e2}}=0.95\)), M5 is formed and the virtual cathode potential oscillates between \(\phi_\text{ref,1}\) and \(\phi_\text{ref,2}\) [Fig.~\ref{fig:Fig1}(a5)].
Then both the low- and high-energy electrons can oscillate at their respective frequencies, generating a combined oscillation waveform [see Fig.~\ref{fig:Fig1}(b5)].
Note that the oscillatory modes exhibit non-sinusoidal waveforms consisting of a dominant frequency $f_\text{os}$ and higher-order harmonics with decreasing amplitudes, where $f_\text{os}$ is related to the electron transit time between the cathode and the reflection point \cite{Lin_2025_pre}.
The phase space distributions of \(\overline{n}_e-E_\text{s}\) for five operating modes (M1--M5) are shown in Figs.~\ref{fig:Fig1}(c1)-(c5), respectively, which confirm that loop trajectories are formed in the oscillation modes.
It should be noted that the finite thermal spread would dampen the space charge oscillations when the effects of non-zero electron temperature are considered \cite{Lafleur_2020}. 
For two electron beams with non-zero thermal temperature, the distributions of electron energy might overlap when the isotropic thermal velocity has a relatively large full width at half maximum, causing the dual-energy beams to degenerate into a continuous distribution with broad energy band, which can potentially alter the determined charge transport mode.

\begin{figure*}[htbp]
\centering
\includegraphics[clip, width=\linewidth]{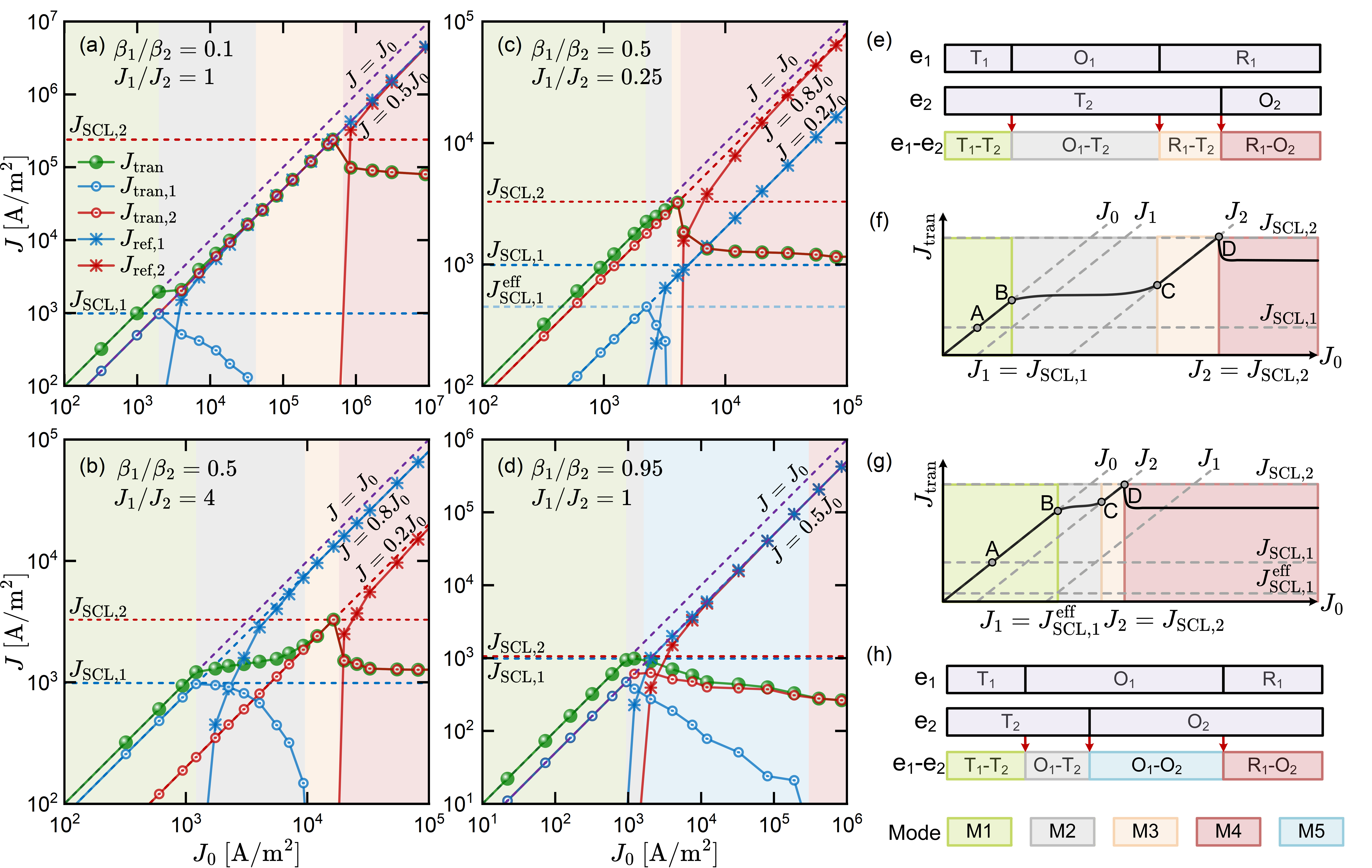}
\caption{\label{fig:Fig3}
Total transmitted current \(J_\text{tran}\) (\(J_\text{tran} = J_\text{tran,1} + J_\text{tran,2}\) with \(J_\text{tran,1}\) for low-energy electron \(\text{e}_1\) and \(J_\text{tran,2}\) for high-energy electron \(\text{e}_2\)) and reflected current \(J_\text{ref}\) (\(J_\text{ref} = J_\text{ref,1} + J_\text{ref,2}\) with \(J_\text{ref,1}\) for \(\text{e}_1\) and \(J_\text{tran,2}\) for \(\text{e}_2\)) as a function of the injected current \(J_0\) (\(J_0 = J_1 + J_2\) with \(J_1\) for \(\text{e}_1\) and \(J_2\) for \(\text{e}_2\)) under different conditions. Governing parameters (\(\beta_1/\beta_2,\ J_1/J_2\))= (0.1, 1) for (a), (0.5, 4) for (b), (0.5, 0.25) for (c), and (0.95, 1) for (d).
(e) Schematic of the operating mode transition for panel (a). (f)--(g) Illustrations of the transmitted current for panels (b) and (c). (h) Schematic of the operating mode transition for panel (d).
}
\end{figure*}
\twocolumngrid

\textit{Transmitted current characteristics}---The transmitted current characteristics for mono-energetic electrons have been explicitly demonstrated in previous studies \coloredcite{Lafleur_2020,Lin_2025_pre}.
The SCL current, as the critical current threshold \coloredcite{Lafleur_2020}, is expressed as 
\begin{equation}
J_\text{SCL}={\frac{4\varepsilon_0}{9}\sqrt{\frac{2e}{m_e}}\frac{V^{3/2}}{d^2}\left(\beta+\sqrt{1+\beta^2}\right)^3},
\end{equation}
where \(\beta\) quantifies the initial velocity of emitted electrons.
The transmitted current \(J_\text{tran}\) equals the injected current \(J_\text{inj}\) for \(J_\text{inj}\leq J_\text{SCL}\) while becomes saturatee for \(J_\text{inj} > J_\text{SCL}\).
However, in dual-energy electron beam driven diodes, each electron beam possesses its own SCL current (\(J_\text{SCL,1}\) for \(\text{e}_1\) and \(J_\text{SCL,2}\) for \(\text{e}_2\)) since \(\beta\) is equal to \(\beta_1\) or \(\beta_2\), respectively, which leads to significantly complex transmitted current behaviors.

By separately tuning the control parameters for dual-energy electron beams, the transmitted current scaling and charge transport mode transition are determined, as shown in Fig.~\ref{fig:Fig3}.
The ratio of initial velocity for two electron beams, \(\beta_1/\beta_2\), plays a significant role in determining the charge transport mode transition.
For electron beams with a large difference in velocity (energy) (e.g., \(\beta_1/\beta_2=0.1\)), the transmitted current characteristics are illustrated in Fig.~\ref{fig:Fig3}(a).
When the injected current \(J_0\) is sufficiently low (below the SCL current of the low-energy electrons \(J_\text{SCL,1}\)), both low- and high-energy electrons can be fully transmitted [mode M1]. 
When the injected current for low-energy electrons \(J_1\) exceeds their corresponding SCL current \(J_\text{SCL,1}\), the transport of low-energy electrons \(\text{e}_1\) is suppressed while the high-energy electrons \(\text{e}_2\) can still be transported completely [mode M2].
Notably, in this regime, the transmitted current for low-energy electrons \(J_\text{tran,1}\) falls below \(J_1\) and the reflected current for low-energy electrons \(J_\text{ref,1}\) increases substantially, maintaining the current conservation \(J_1=J_\text{tran,1}+J_\text{ref,1}\).
As the injected current further increases, low-energy electrons can be fully reflected with \(J_\text{ref,1}=J_\text{inj,1}\), while high-energy electrons remain fully transported [mode M3].
Once the injected current for high-energy electrons \(J_2\) is larger than their corresponding SCL current \(J_\text{SCL,2}\), high-energy electrons \(\text{e}_2\) are partially reflected, accompanied with the entire reflection of \(\text{e}_1\) [mode M4].
Consequently, while increasing the total injected current \(J_0\), the charge transport modes undergo a well-defined pathway: \(\text{M}1\rightarrow\text{M}2\rightarrow\text{M}3\rightarrow\text{M}4\), which can be abstracted to discrete eigen-states defined by combinatorial permutations of the transport modes for both energy components [as illustrated later in Fig.~\ref{fig:Fig3}(e)].

Figure~\ref{fig:Fig3}(b) maps the transmitted current characteristics under parameter conditions of \(\beta_1/\beta_2=0.5\) and \(J_1/J_2=4\), demonstrating more universal behaviors of the transmitted currents.
Note that the critical mode transitions are quantitatively defined by the intersection points where the injected current of each electron beam component reaches its corresponding SCL current.
As illustrated in the schematic diagram [see Fig.~\ref{fig:Fig3}(f)], we identify several critical points (A-D) to visually mark the evolution of transmitted current.
At point A, the total injected current \(J_0=J_\text{SCL,1}\) and all the electrons can be transported steadily, i.e., M1 with the total transmitted current \(J_\text{tran}=J_0\).
The transition from M1 to M2 occurs abruptly when the injected current of low-energy electrons reaches their SCL current [point B], i.e., \(J_1=J_\text{SCL,1}\), and then \(J_\text{tran}\) deviates from \(J_0\).
Theoretically, the transition point is expected when \(J_1<J_\text{SCL,1}\); however, due to the significant energy difference between the two electron beams and the relatively minor space charge effect of the high-energy electrons, one can consider the critical transition at point B.
The point C marks the M2--M3 transition where \(J_\text{tran}=J_2\) and \(J_\text{tran,1}=0\), indicating that only the single electron beam (\(\text{e}_2\)) is transmitted in the diode.
At point D, \(J_2=J_\text{SCL,2}\) and the transmitted current achieves its maximum, simultaneously signifying the M3--M4 transition.

In addition, the injected current ratio \(J_1/J_2\) serves as a key control parameter, modulating the transmitted current characteristics.
Figure~\ref{fig:Fig3}(c) shows that under high-energy electron-dominated injection conditions (\(J_1/J_2=0.25<1\)), the transmitted current characteristics exhibits an obvious difference from that in low-energy electron-dominated injection conditions (\(J_1/J_2=4>1\)) [see Fig.~\ref{fig:Fig3}(b)].
Notably, low-energy electrons \(\text{e}_1\) undergo abnormal reflection before reaching their conventional SCL current \(J_\text{SCL,1}\), showing an effective SCL current \(J_\text{SCL,1}^\text{eff}\) smaller than \(J_\text{SCL,1}\).
As illustrated in Fig.~\ref{fig:Fig3}(g), the critical point B for M1--M2 transition is defined as the intersection point with \(J_1=J_\text{SCL,1}^\text{eff}\), which occurs before reaching the conventional threshold \(J_1=J_\text{SCL,1}\).
The virtual cathode, primarily sustained by the high-energy electrons, preferentially reflects the low-energy electrons due to their lower reflection threshold, even while low-energy electron injected current remains below the conventional SCL current, resulting in effective SCL current \(J_\text{SCL,1}^\text{eff}<J_\text{SCL,1}\). 
In this regime, the low-energy electron transport is extrinsically suppressed by the space charge effects from the high-energy electrons rather than being limited by its own space charge effects.
This discovery reveals that the SCL threshold for each electron beam in dual-energy electron beam diodes is cross-modulated and mutually coupled by two electron beam parameters, which is distinct from the single-energy electron beam diodes where the SCL current is solely determined by the electron beam parameters (electron energy) and gap properties (gap distance and applied voltage).

The charge transport mode transition when the two electron beams have nearly identical energies (e.g., \(\beta_1/\beta_2=0.95\)) is shown in Fig.~\ref{fig:Fig3}(d).
Since the energy difference between the two electron beams (\(\text{e}_1\) and \(\text{e}_2\)) is negligible, their combined behavior in terms of charge transport closely resembles that of a single beam.
Practically, the two groups of electrons can be treated as equivalent when determining the transmitted current. Therefore, the premature reflection of each beam (\(\text{e}_1\) or \(\text{e}_2\)) before reaching its respective SCL current is physically reasonable.
Neither the \(\text{e}_1-\)beam nor the \(\text{e}_2-\) beam individually reaches its own SCL threshold, but the total injected current from both does.
Once the total injected current \(J_0\) reaches the SCL threshold, transport for both beams is simultaneously suppressed via partial reflection [M5, also see Fig.~\ref{fig:Fig1}(a5)].
Note that the beam electrons at lower energy are more strongly reflected when the total current is sufficiently large [see Fig.~\ref{fig:Fig3}(d)].
Figure~\ref{fig:Fig3}(h) presents the charge transport mode transition pathway (\(\text{M1}\rightarrow\text{M2}\rightarrow\text{M5}\rightarrow\text{M4}\)) for \(\beta_1/\beta_2\rightarrow1\), which differs from the observed pathway for \(\beta_1/\beta_2<1\) [see Fig.~\ref{fig:Fig1}(e)].

\begin{figure}[htbp]
\centering
\includegraphics[clip, width=0.9\linewidth]{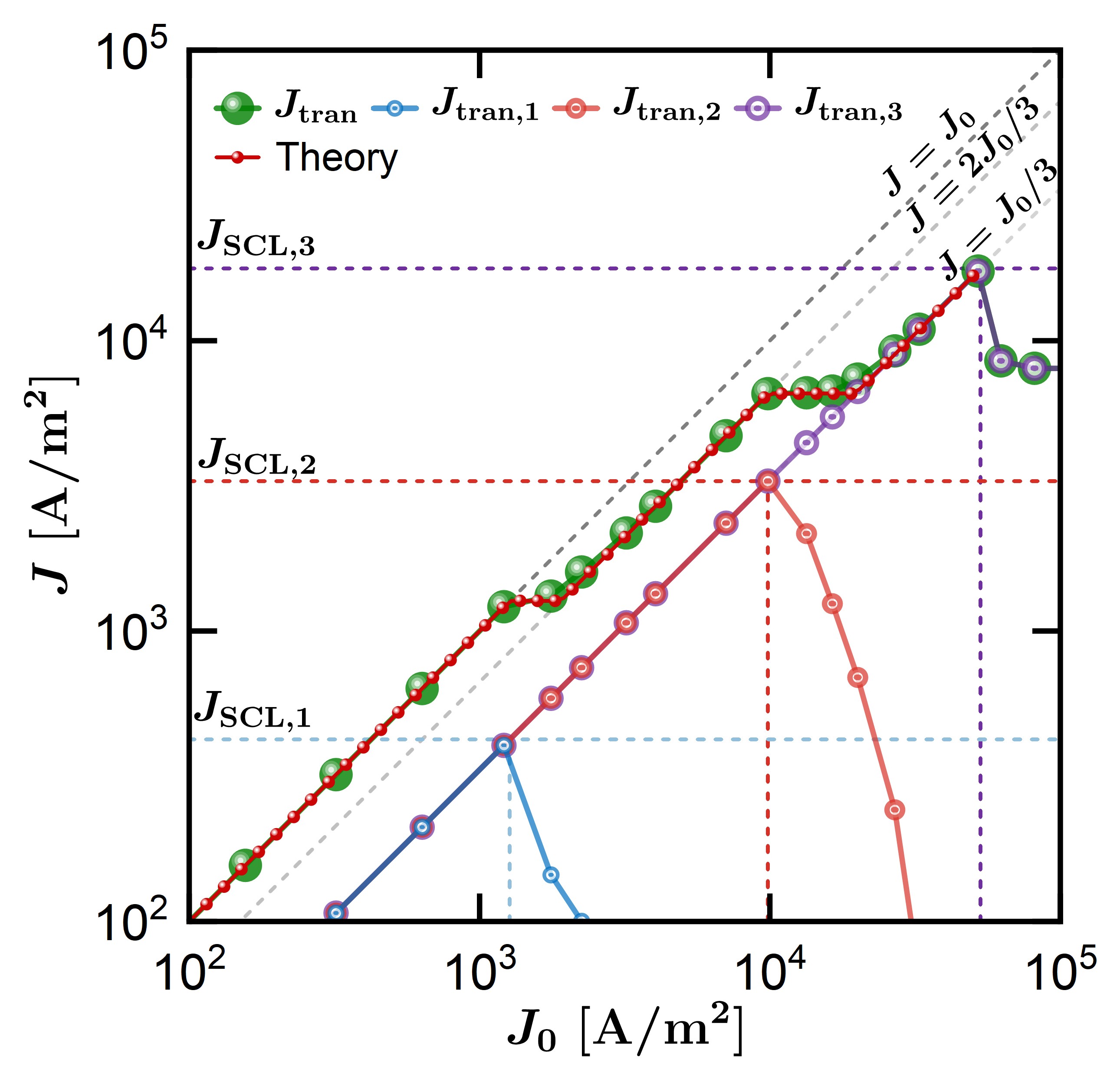}
\caption{\label{fig:Fig4} Transmitted current density \(J_\text{tran}\) versus the total injected current density \(J_0\). The prediction from the proposed theoretical formula [Eq.~\eqref{EQ3}] agrees well with PIC simulation results, for example, with the electron beam number \(n = 3\). 
}
\end{figure}

A more generalized analysis can be conducted for \(n\)--component electron beams, namely \(\{\)\(\text{e}_1\), \(\text{e}_2\), ..., \(\text{e}_n\)\(\}\) with energies at \(\mathcal{E}_{e1}<\mathcal{E}_{e2}<...<\mathcal{E}_{en}\).
Each electron beam \(\text{e}_i\) (\(1\leq i \leq n\)) exhibits three possible transport states \(S_i\in\{2,\ 1,\ 0\}\), in which \(S_i=2\) for T--mode, \(S_i=1\) for O--mode, and \(S_i=0\) for R--mode, respectively.
The operating modes in diodes with \(n\)--component electron beams can be labeled as \((S_1S_2...S_n)\), yielding \(3^n\) possible combinations.
However, the space-charge effect preferentially suppresses the transport of lower-energy electrons, resulting in \(S_n\geq ...\geq S_2 \geq S_1\), and thus the number of operating modes is determined as \(\text{C}_{n+2}^2\).
Since one specific operating mode with all the electron beams simultaneously reflected is physically inadmissible \([(S_1S_2...S_n)\neq (00...0)]\), the number of potential operating modes with \(n\)--component electron beams is determined by
\begin{equation}
N_\text{mode}=\text{C}_{n+2}^2-1={n(n+3)}/2.
\end{equation}
Here, five operating modes \((N_\text{mode}=5)\) are determined for \(n=2\). 
If one considers identical injected current for \(n\)--component electron beams, \(J_1=J_2=...=J_n=J_0/n\), with substantial energy differences.
The transmitted current \(J_\text{tran}\) versus the total injected current \(J_0\) exhibits a stepwise growth [see Fig.~\ref{fig:Fig4}], and the transmitted current for \(\text{e}_i\) reaches its maximum \(J_{\text{tran},i}=J_0/n=J_{\text{SCL},i}\) when \(J_i=J_{\text{SCL},i}\).
If \(J_i>J_{\text{SCL},i}\) for \(\text{e}_i\), the transmitted component progressively decreases to zero, and the mode transition \(\text{T}\rightarrow\text{O}\rightarrow\text{R}\) occurs.
Generally, with an increasing \(J_0\), the intensified space-charge effects lead to complete reflections of electron beams \(\text{e}_i\) in the energy sequence, showing stepwise growth in transmitted current. 
The total transmitted current can be described as
\begin{equation}
J_\text{tran}(J_0)=\left\{
\begin{aligned}
&\frac{n+1-i}{n}J_0,\ J_0\in\left[J_\text{a},\ J_\text{b}\right),\\
&(n+1-i)J_{\text{SCL},i},\ J_0\in\left[J_\text{b},\ J_\text{c}\right),
\end{aligned}
\right.
\label{EQ3}
\end{equation}
where \(J_\text{a}={(n+2-i)nJ_{\text{SCL},i-1}}/{(n+1-i)}\), \(J_\text{b}=nJ_{\text{SCL},i}\), \(J_\text{c}={(n+1-i)nJ_{\text{SCL},i}}/{(n-i)}\), and \(i\) varies from 1 to \(n\).
The maximum transmitted current \(J_\text{tran}^\text{max}=J_{\text{SCL},n}\), independent of the injected current \(J_0\), is reached when the highest-energy electron beam \(\text{e}_n\) reaches its SCL current, i.e., \(J_n=J_{\text{SCL},n}\).

\textit{Conclusion}---In conclusion, we explicitly determined five complete modes for charge transport in dual-energy electron beam driven diodes, which are modulated by controlling the two electron beam parameters.
The initial velocity ratio and the injected current ratio are identified as governing parameters; the former determines the charge transport modes and their transitions, while the latter modulates the characteristics of the transmitted current.
The space charge limited current for each electron beam in dual-energy electron beam diodes is no longer an intrinsic property but emerges as a coupled threshold cross-modulated by two electron beams.
The discovery provides a fundamental advance in multi-energy beam physics, enabling precise tuning strategies for electron beam-driven diodes, especially for applications in modern vacuum electronic devices.


\vspace{2 mm}
\textit{Acknowledgments}---The authors acknowledge the financial support from the Fundamental and Interdisciplinary Disciplines Breakthrough Plan of the Ministry of Education of China (Grant No.~JYB2025XDXM312) and the Beijing Natural Science Foundation (Grant No.~3244040).

\vspace{2 mm}
\textit{Data availability}---The data that support the findings of this study are available from the authors upon reasonable request. 

\Large
\bibliography{references}

\end{document}